\documentclass[aps,prd,onecolumn,nofootinbib,superscriptaddress]{revtex4-2}

\usepackage[T1]{fontenc}
\usepackage[utf8]{inputenc}
\usepackage{lmodern}
\usepackage{amsmath,amssymb,amsthm,mathtools,bm}
\usepackage{physics}
\usepackage{hyperref}
\hypersetup{colorlinks=true,citecolor=blue,linkcolor=blue,urlcolor=blue}

\newcommand{\Mpl}{M_{\rm Pl}}

\begin{document}

\title{The Wave Function of the Universe and Inflation}

\author{Gerasimos Kouniatalis}
\email{gkouniatalis@noa.gr}
\affiliation{Physics Department, National Technical University of Athens, 15780 Zografou Campus, Athens, Greece}
\affiliation{National Observatory of Athens, Lofos Nymfon, 11852 Athens, Greece}

\begin{abstract}
We develop a minimal framework that connects the wave function of the universe to inflation in closed Friedmann-Robertson-Walker minisuperspace with a homogeneous scalar field. Starting from the minisuperspace action, we derive the Wheeler-DeWitt equation, its Hamilton-Jacobi limit, and the exact identifications between the gradients of the WKB phase and the classical background quantities that govern the cosmological evolution. In the inflationary regime, where spatial curvature has been diluted, this formulation yields direct expressions for the Hubble slow-roll parameter, the tensor-to-scalar ratio, the scalar amplitude, the scalar spectral tilt, and the inflationary energy scale in terms of the physical phase gradients of the wave function.
We then evaluate the Euclidean saddle for slowly varying scalar configurations and derive the corresponding semiclassical branch weights associated with the standard no-boundary and tunneling prescriptions. This establishes a clean separation between the dynamical information encoded in the WKB phase and the probabilistic weighting of inflationary histories encoded in the boundary condition. Within this framework, we obtain a measure-independent threshold weight for achieving a prescribed number of e-folds and, in the plateau regime, express the branch weights directly in terms of inflationary observables. We further translate current cosmic microwave background constraints into direct bounds on the physical phase gradients of the wave function. The resulting framework provides a compact and testable link between quantum cosmology and inflationary phenomenology.
\end{abstract}

\maketitle

\section{Introduction}

Inflation provides the standard framework for understanding the observed large-scale homogeneity of the universe, its near-spatial flatness, and the origin of primordial perturbations with an approximately scale-invariant spectrum \cite{Starobinsky1980,Guth1981,Linde1982,LythRiotto1999,PlanckInflation2018,Baumann2009Inflation,Baumann2018Primordial,BaumannMcAllister2015,Durrer2004Perturbations,Durrer2001CMB,Durrer1996CMB,Riotto2003Perturbations}. In the conventional effective-field-theory treatment, one specifies a scalar potential and derives the corresponding cosmological background evolution together with the spectrum of perturbations generated during the accelerated expansion \cite{Durrer2008CMBBook,BurgessLeeTrott2009,BurgessHolmanTasinatoWilliams2015,RenauxPetel2015,RenauxPetelMizunoKoyama2011,BaumannGreenLeePorto2016,DeSimoneRiotto2013,DeSimonePerrierRiotto2013,GermaniKehagias2010PRL,GermaniKehagias2010JCAP,GermaniKehagias2011,KouniatalisSaridakis2025GEP}. Quantum cosmology addresses the deeper question of how such inflationary histories emerge at the level of the wave function of the universe and how that wave function constrains the initial conditions for an inflationary spacetime \cite{HartleHawking1983,Vilenkin1982,Vilenkin1984,Halliwell1988,Lehners2023,PartoucheToumbasDeVaulchier2021,PartoucheToumbasDeVaulchier2021BSM,KaimakkamisSil2024,KaimakkamisPartoucheSilToumbas2025,Rondeau2024,BasilakosKouniatalisSaridakisTzerefos2026}.

A natural setting for this question is the minisuperspace description of a homogeneous and isotropic universe coupled to a homogeneous scalar field. In that setting, the gravitational and matter degrees of freedom are reduced to a finite-dimensional configuration space, and the corresponding quantum dynamics is encoded in the Wheeler-DeWitt equation. Its semiclassical solutions define WKB branches on which the phase of the wave function plays the role of a Hamilton principal function. The gradients of this phase then determine the canonical momenta associated with the background variables and thereby supply the quantities that govern the classical cosmological evolution. This provides a direct and physically transparent relation between the semiclassical structure of the wave function and the inflationary background.

A central theme of the present work is the separation between two distinct pieces of information contained in the quantum-cosmological description. The first is the dynamical information carried by the WKB phase, which determines the classical inflationary branch and its kinematics. The second is the probabilistic information associated with the choice of boundary condition for the wave function, which assigns relative weights to different semiclassical branches. When these two roles are kept conceptually distinct, the connection between quantum cosmology and inflationary phenomenology becomes especially transparent. The minisuperspace potential specifies the underlying model, the semiclassical phase determines the corresponding classical trajectory, and the boundary condition selects the relative weighting of admissible inflationary histories.

This viewpoint leads to a compact framework in which the wave function of the universe can be connected directly to observables relevant for inflation. In the regime where spatial curvature has already been diluted and the background is in the inflationary attractor, the physical phase gradients associated with the semiclassical branch determine the quantities that control the standard slow-roll observables. As a result, the scalar amplitude, the tensor-to-scalar ratio, the scalar tilt, and the characteristic inflationary energy scale can be expressed directly in terms of the phase-space data extracted from the wave function. This gives a direct phenomenological interpretation to the semiclassical branch structure of the Wheeler-DeWitt solution \cite{MukhanovChibisov1981,Hawking1982,StarobinskyPert1982,GuthPi1982,Bardeen1983,MukhanovReview1992,StewartLyth1993}.

The same framework also admits a controlled description of the nucleation stage through a compact Euclidean saddle for slowly varying scalar configurations. Evaluating this saddle yields the corresponding semiclassical branch weights for the standard no-boundary and tunneling prescriptions. In this way, the boundary condition enters the formalism through a well-defined weighting of inflationary initial data, while the subsequent classical evolution remains governed by the WKB phase. This separation makes it possible to formulate model-independent statements about the probability of obtaining a given amount of inflation and to translate observational constraints from the cosmic microwave background into direct bounds on the physical phase gradients of the wave function.

The purpose of this paper is to develop this connection systematically in the closed Friedmann-Robertson-Walker minisuperspace with a homogeneous scalar field. We derive the reduced action, the Hamiltonian constraint, and the corresponding Wheeler-DeWitt equation, and then analyze its semiclassical limit in a form suited for inflationary cosmology. We show how the classical observables are recovered from the phase of the wave function, derive the Euclidean saddle and the associated branch weights, and finally combine these ingredients into a phenomenological framework that links quantum-cosmological initial conditions to inflationary observables in a direct and testable way.

The paper is organized as follows. Section~\ref{sec:minisuperspace} reviews the closed minisuperspace action, the canonical structure, and the Wheeler-DeWitt equation. Section~\ref{sec:phase} develops the extraction of classical observables from the WKB phase. Section~\ref{sec:weights} derives the Euclidean saddle action and the corresponding semiclassical branch weights. Section~\ref{sec:phenomenology} translates these results into phenomenological statements, including direct bounds on the phase gradients from current cosmic microwave background data and threshold weights for obtaining a prescribed duration of inflation. Section~\ref{conclusions} summarizes the main contributions of the present work.

\section{Closed minisuperspace and the Wheeler-DeWitt equation}
\label{sec:minisuperspace}

\subsection{Action and conventions}

We consider Einstein gravity minimally coupled to a canonical scalar field with potential $V(\phi)$, described by the four-dimensional action
\begin{equation}
S = \int d^4x\,\sqrt{-g}\left[
\frac{\Mpl^2}{2}R
-\frac12 g^{\mu\nu}\partial_\mu\phi\,\partial_\nu\phi
- V(\phi)
\right],
\label{eq:4d-action}
\end{equation}
where the reduced Planck mass is defined by
\begin{equation}
\Mpl \equiv (8\pi G)^{-1/2},
\end{equation}
and we work in units with $\hbar=c=1$.

In order to describe both the compact nucleation saddle and the subsequent Lorentzian inflationary evolution within a single minisuperspace framework, we adopt the closed Friedmann-Robertson-Walker ansatz
\begin{equation}
ds^2 = -N(t)^2 dt^2 + a(t)^2 d\Omega_3^2,
\qquad
a(t)=e^{\alpha(t)},
\label{eq:closed-frw}
\end{equation}
where $N(t)$ is the lapse function, $a(t)$ is the scale factor, and $d\Omega_3^2$ denotes the metric on the unit three-sphere. The volume of the unit three-sphere is
\begin{equation}
\mathrm{Vol}(S^3)=2\pi^2.
\end{equation}
The scalar field is taken to be homogeneous,
\begin{equation}
\phi = \phi(t).
\end{equation}

Substituting the metric ansatz \eqref{eq:closed-frw} into the action \eqref{eq:4d-action}, integrating over the spatial three-sphere, and discarding the standard total derivative term yields the reduced minisuperspace action
\begin{equation}
S = 2\pi^2 \int dt\, e^{3\alpha}
\left[
-3\Mpl^2 \frac{\dot\alpha^2}{N}
+ 3 \Mpl^2 N e^{-2\alpha}
+ \frac{\dot\phi^2}{2N}
- N V(\phi)
\right].
\label{eq:minisup-action}
\end{equation}
This reduced action makes the physical content of the model fully explicit. The variable $\alpha=\ln a$ represents the logarithm of the scale factor, the lapse $N$ enforces time-reparametrization invariance, the term proportional to $\dot{\alpha}^2$ is the gravitational kinetic contribution, the term proportional to $e^{-2\alpha}$ encodes the positive spatial curvature of the closed universe, the term proportional to $\dot{\phi}^2$ is the scalar kinetic energy, and the final term is the scalar potential energy.

\subsection{Canonical momenta and Hamiltonian constraint}

The canonical momenta derived from \eqref{eq:minisup-action} are
\begin{align}
p_\alpha &\equiv \pdv{L}{\dot\alpha}
= -12\pi^2 \Mpl^2 e^{3\alpha}\frac{\dot\alpha}{N},
\label{eq:palpha}\\[1mm]
p_\phi &\equiv \pdv{L}{\dot\phi}
= 2\pi^2 e^{3\alpha}\frac{\dot\phi}{N}.
\label{eq:pphi}
\end{align}
Solving these relations for the velocities gives
\begin{align}
\dot\alpha &= - \frac{N}{12\pi^2 \Mpl^2} e^{-3\alpha} p_\alpha,
\label{eq:adot-from-p}\\[1mm]
\dot\phi &= \frac{N}{2\pi^2} e^{-3\alpha} p_\phi.
\label{eq:phidot-from-p}
\end{align}

The Legendre transform then leads to the Hamiltonian
\begin{equation}
H = p_\alpha \dot\alpha + p_\phi \dot\phi - L
= N \,\mathcal{H},
\end{equation}
with Hamiltonian constraint
\begin{equation}
\mathcal{H}
=
e^{-3\alpha}
\left[
-\frac{p_\alpha^2}{24\pi^2 \Mpl^2}
+\frac{p_\phi^2}{4\pi^2}
-6\pi^2 \Mpl^2 e^{4\alpha}
+2\pi^2 e^{6\alpha}V(\phi)
\right].
\label{eq:Hconstraint}
\end{equation}
Because the lapse function acts as a Lagrange multiplier, the classical dynamics is subject to the constraint
\begin{equation}
\mathcal{H}=0.
\label{eq:Hzero}
\end{equation}

This constraint reproduces the standard Friedmann equation for a closed universe. After variation, one may set $N=1$. Substituting \eqref{eq:palpha} and \eqref{eq:pphi} into \eqref{eq:Hzero} yields
\begin{equation}
-6\pi^2 \Mpl^2 e^{3\alpha}\dot\alpha^2
+\pi^2 e^{3\alpha}\dot\phi^2
-6\pi^2 \Mpl^2 e^{\alpha}
+2\pi^2 e^{3\alpha}V(\phi)=0.
\end{equation}
Dividing by $2\pi^2 e^{3\alpha}$ gives
\begin{equation}
-3\Mpl^2 \dot\alpha^2
+\frac12 \dot\phi^2
-3\Mpl^2 e^{-2\alpha}
+V(\phi)=0.
\end{equation}
Since the Hubble parameter is $H\equiv \dot a/a = \dot\alpha$, this relation becomes
\begin{equation}
3\Mpl^2\left(H^2 + e^{-2\alpha}\right)
=
\frac12 \dot\phi^2 + V(\phi).
\label{eq:closed-friedmann}
\end{equation}
Equation \eqref{eq:closed-friedmann} is precisely the standard Friedmann equation for a spatially closed Friedmann-Robertson-Walker universe.

\subsection{Quantization}

Canonical quantization promotes the momenta to differential operators according to
\begin{equation}
p_\alpha \rightarrow -i\partial_\alpha,
\qquad
p_\phi \rightarrow -i\partial_\phi,
\end{equation}
so that the Hamiltonian constraint becomes the Wheeler-DeWitt equation
\begin{equation}
\left[
\frac{1}{24\pi^2 \Mpl^2}\partial_\alpha^2
-
\frac{1}{4\pi^2}\partial_\phi^2
-
6\pi^2 \Mpl^2 e^{4\alpha}
+
2\pi^2 e^{6\alpha}V(\phi)
\right]\Psi(\alpha,\phi)=0,
\label{eq:WDW}
\end{equation}
where we have adopted the simplest factor ordering. The precise factor-ordering choice is not essential for the semiclassical analysis developed below.

The kinetic structure implicit in \eqref{eq:WDW} endows minisuperspace with a Lorentzian signature. As a result, the Wheeler-DeWitt equation is hyperbolic rather than elliptic. This feature is crucial in the semiclassical regime, where it permits one of the minisuperspace variables to play the role of an emergent time parameter along a classical branch.

\subsection{WKB classicality}

To analyze the semiclassical regime, we write the wave function in polar form,
\begin{equation}
\Psi(\alpha,\phi)=A(\alpha,\phi)\,e^{iS(\alpha,\phi)}.
\label{eq:wkb}
\end{equation}
Substituting \eqref{eq:wkb} into \eqref{eq:WDW} and separating real and imaginary parts yields exact equations involving derivatives of both the amplitude $A$ and the phase $S$. In the WKB regime, the phase varies parametrically more rapidly than the amplitude, so that
\begin{equation}
\abs{\partial_A S} \gg \abs{\partial_A \ln A},
\qquad
A\in\{\alpha,\phi\},
\label{eq:classicality}
\end{equation}
and the leading real part reduces to the Hamilton-Jacobi equation
\begin{equation}
-\frac{1}{24\pi^2 \Mpl^2}\left(S_{,\alpha}\right)^2
+\frac{1}{4\pi^2}\left(S_{,\phi}\right)^2
-6\pi^2 \Mpl^2 e^{4\alpha}
+2\pi^2 e^{6\alpha}V(\phi)
=0.
\label{eq:HJ}
\end{equation}
Here and in what follows we use the notation
\begin{equation}
S_{,\alpha}\equiv \partial_\alpha S,
\qquad
S_{,\phi}\equiv \partial_\phi S.
\end{equation}

At the same order, the imaginary part gives the conserved current equation
\begin{equation}
\partial_\alpha\!\left(
\frac{A^2}{12\pi^2 \Mpl^2} S_{,\alpha}
\right)
-
\partial_\phi\!\left(
\frac{A^2}{2\pi^2}S_{,\phi}
\right)
=0,
\label{eq:current}
\end{equation}
which is the minisuperspace analogue of the standard WKB continuity equation.

Equation \eqref{eq:HJ} admits the usual Hamilton-Jacobi interpretation, with the phase gradients identified as classical canonical momenta:
\begin{equation}
p_\alpha = S_{,\alpha},
\qquad
p_\phi = S_{,\phi}.
\label{eq:momenta-from-S}
\end{equation}
Combining \eqref{eq:momenta-from-S} with \eqref{eq:adot-from-p} and \eqref{eq:phidot-from-p} shows that the classical evolution along a semiclassical branch is determined directly by the phase of the wave function. In this sense, the WKB phase supplies the bridge between the Wheeler-DeWitt description and the classical cosmological observables that characterize the inflationary background.
\section{Inflationary observables from the WKB phase}
\label{sec:phase}

\subsection{Physical phase gradients}

The canonical momenta $p_\alpha$ and $p_\phi$ scale extensively with the physical volume $a^3=e^{3\alpha}$. It is therefore natural to introduce the corresponding volume-rescaled quantities
\begin{equation}
\Pi_\alpha \equiv \frac{e^{-3\alpha}}{2\pi^2} S_{,\alpha},
\qquad
\Pi_\phi \equiv \frac{e^{-3\alpha}}{2\pi^2} S_{,\phi},
\label{eq:Pi-def}
\end{equation}
which represent the physically relevant phase gradients along a semiclassical branch. Using the Hamilton-Jacobi identifications \eqref{eq:momenta-from-S}, together with the classical relations \eqref{eq:adot-from-p} and \eqref{eq:phidot-from-p}, one obtains
\begin{align}
H \equiv \dot\alpha
&= -\frac{1}{12\pi^2 \Mpl^2} e^{-3\alpha}S_{,\alpha}
= -\frac{\Pi_\alpha}{6\Mpl^2},
\label{eq:H-from-S}\\[1mm]
\dot\phi
&= \frac{1}{2\pi^2} e^{-3\alpha}S_{,\phi}
= \Pi_\phi.
\label{eq:phidot-from-S}
\end{align}
Accordingly, the physical interpretation of the rescaled phase gradients is
\begin{equation}
\Pi_\alpha = -6\Mpl^2 H,
\qquad
\Pi_\phi = \dot\phi.
\end{equation}
These relations provide the direct link between the WKB phase of the wave function and the background kinematics of inflationary evolution. In particular, the geometric phase gradient determines the Hubble expansion rate, while the scalar phase gradient determines the field velocity along the classical branch.

\subsection{Background relations during observable inflation}

The inflationary observables relevant for the cosmic microwave background are evaluated well after the initial curvature contribution of a closed universe has been exponentially suppressed. It is therefore appropriate, for the remainder of this section, to work in the regime
\begin{equation}
e^{-2\alpha}\ll H^2,
\label{eq:curvature-negligible}
\end{equation}
for which the background dynamics is effectively that of a spatially flat universe.

For a canonically normalized scalar field in this regime, the first Hubble slow-roll parameter is
\begin{equation}
\epsilon_1 \equiv -\frac{\dot H}{H^2}
= \frac{\dot\phi^2}{2\Mpl^2 H^2}.
\label{eq:epsilon-def}
\end{equation}
Substituting \eqref{eq:H-from-S} and \eqref{eq:phidot-from-S} gives
\begin{equation}
\epsilon_1
=
18\,\Mpl^2\left(\frac{\Pi_\phi}{\Pi_\alpha}\right)^2.
\label{eq:epsilon-from-S}
\end{equation}
This expression shows that the slow-roll condition is encoded directly in the relative magnitude of the scalar and geometric phase gradients. In particular, the slow-roll regime corresponds to
\begin{equation}
\abs{\Pi_\phi} \ll \frac{\abs{\Pi_\alpha}}{\sqrt{18}\,\Mpl}.
\end{equation}

It is also useful to express the evolution along the classical trajectory in terms of the number of e-folds, $N\equiv \ln a$. Since $dN/dt=H$, one finds
\begin{equation}
\frac{d\phi}{dN}
=
\frac{\dot\phi}{H}
=
-6\Mpl^2 \frac{\Pi_\phi}{\Pi_\alpha}.
\label{eq:dphidN}
\end{equation}
More generally, for any function $F(\alpha,\phi)$ on minisuperspace, differentiation along the classical branch yields
\begin{equation}
\frac{dF}{dN}
=
\partial_\alpha F
-6\Mpl^2 \frac{\Pi_\phi}{\Pi_\alpha}\partial_\phi F.
\label{eq:dFdN}
\end{equation}
It is convenient to define the corresponding derivative operator
\begin{equation}
D_N \equiv
\partial_\alpha
-6\Mpl^2 \frac{\Pi_\phi}{\Pi_\alpha}\partial_\phi.
\label{eq:DN}
\end{equation}
The second Hubble slow-roll parameter may then be written as
\begin{equation}
\epsilon_2 \equiv \frac{d\ln \epsilon_1}{dN}
= D_N \ln \epsilon_1.
\label{eq:epsilon2}
\end{equation}
The relations above establish that the background hierarchy relevant for inflation can be expressed entirely in terms of the directional properties of the WKB phase.

\subsection{Scalar and tensor observables}

At first order in slow roll, the standard expressions for the scalar amplitude, tensor-to-scalar ratio, and scalar spectral tilt are \cite{MukhanovReview1992,StewartLyth1993,PlanckInflation2018}
\begin{equation}
A_s = \frac{H^2}{8\pi^2 \Mpl^2 \epsilon_1},
\qquad
r = 16\epsilon_1,
\qquad
n_s-1 = -2\epsilon_1 - \epsilon_2.
\label{eq:standard-obs}
\end{equation}
Using \eqref{eq:H-from-S}, \eqref{eq:epsilon-from-S}, and \eqref{eq:epsilon2}, these observables can be written directly in terms of the WKB phase gradients.

For the tensor-to-scalar ratio one obtains
\begin{equation}
r
=
288\,\Mpl^2\left(\frac{\Pi_\phi}{\Pi_\alpha}\right)^2.
\label{eq:r-from-S}
\end{equation}

\begin{figure}[t]
\centering
\includegraphics[width=0.82\textwidth]{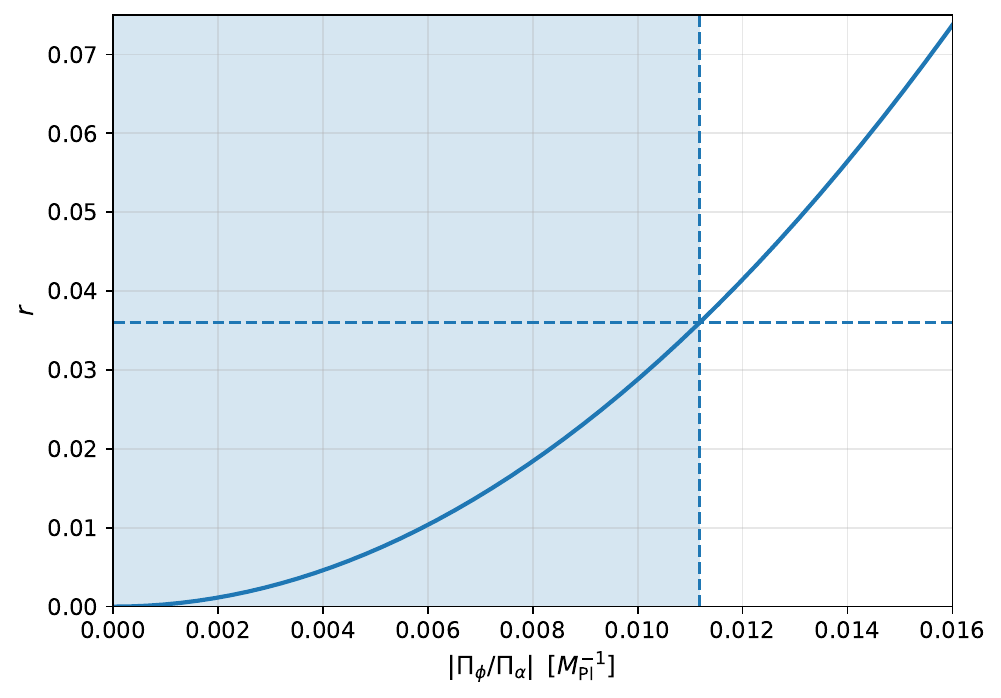}
\caption{Tensor-to-scalar ratio $r$ as a function of the magnitude of the ratio of physical WKB phase gradients, $|\Pi_\phi/\Pi_\alpha|$. The solid curve shows the relation
\eqref{eq:r-from-S}.
The shaded region denotes the observationally allowed domain implied by the current upper bound $r_{0.05}<0.036$ at 95\% confidence, and the dashed lines indicate the corresponding limiting values
$r=0.036$
and
$|\Pi_\phi/\Pi_\alpha| \simeq 0.0112\,M_{\rm Pl}^{-1}$.}
\label{fig:r_phase_ratio}
\end{figure}

Figure~\ref{fig:r_phase_ratio} shows the direct relation between the tensor-to-scalar ratio and the magnitude of the ratio of the physical WKB phase gradients. Since the observable prediction is \eqref{eq:r-from-S},
current observational constraints on $r$ translate immediately into an allowed range for $|\Pi_\phi/\Pi_\alpha|$. The shaded region indicates the part of parameter space compatible with the present upper bound, while the dashed horizontal and vertical lines mark the corresponding observational thresholds. The figure makes explicit that the current limit on primordial tensor modes imposes a tight upper bound on the scalar-to-geometric phase-gradient ratio along the semiclassical inflationary branch.

For the scalar amplitude one finds
\begin{equation}
A_s
=
\frac{\Pi_\alpha^4}{5184\pi^2 \Mpl^8 \Pi_\phi^2}.
\label{eq:As-from-S}
\end{equation}
The product $A_s r$ then simplifies to
\begin{equation}
A_s r
=
\frac{\Pi_\alpha^2}{18\pi^2 \Mpl^6}
=
\frac{2H^2}{\pi^2 \Mpl^2}.
\label{eq:Asr-from-S}
\end{equation}
This combination is especially useful because it determines the inflationary Hubble scale directly from observables:
\begin{equation}
H
=
\pi \Mpl \sqrt{\frac{A_s r}{2}}.
\label{eq:H-from-Asr}
\end{equation}

For the scalar tilt one obtains
\begin{equation}
n_s-1
=
-36\,\Mpl^2\left(\frac{\Pi_\phi}{\Pi_\alpha}\right)^2
-
2\,D_N\ln\left|\frac{\Pi_\phi}{\Pi_\alpha}\right|.
\label{eq:ns-from-S}
\end{equation}
This form makes transparent the two distinct contributions to the tilt: a contribution determined by the local magnitude of the ratio of scalar to geometric phase gradients, and a contribution determined by the variation of that ratio along the classical inflationary trajectory.

Since the common prefactor in \eqref{eq:Pi-def} cancels in the ratio, the observable relations may also be written directly in terms of derivatives of the phase itself. In particular,
\begin{equation}
r
=
288\,\Mpl^2
\left(\frac{S_{,\phi}}{S_{,\alpha}}\right)^2,
\label{eq:r-direct-S}
\end{equation}
and similarly
\begin{equation}
n_s-1
=
-36\,\Mpl^2
\left(\frac{S_{,\phi}}{S_{,\alpha}}\right)^2
-
2\,D_N\ln\left|\frac{S_{,\phi}}{S_{,\alpha}}\right|.
\label{eq:ns-direct-S}
\end{equation}
The inflationary observables therefore depend only on the local directional structure of the WKB phase in minisuperspace, rather than on its overall normalization.

\subsection{Potential energy and inflationary scale}

In the effectively flat regime, the Friedmann equation reduces to
\begin{equation}
3\Mpl^2 H^2
=
\frac12 \dot\phi^2 + V(\phi).
\label{eq:flat-friedmann}
\end{equation}
Substituting \eqref{eq:H-from-S} and \eqref{eq:phidot-from-S} gives
\begin{equation}
V(\phi)
=
\frac{\Pi_\alpha^2}{12\Mpl^2}
-
\frac{\Pi_\phi^2}{2}.
\label{eq:V-from-S}
\end{equation}
In the slow-roll regime, the kinetic contribution is subleading, and one obtains the leading approximation
\begin{equation}
V \simeq \frac{\Pi_\alpha^2}{12\Mpl^2}.
\label{eq:V-approx}
\end{equation}
Combining this relation with \eqref{eq:Asr-from-S} yields the familiar observable expression
\begin{equation}
V \simeq \frac{3\pi^2}{2}\,\Mpl^4\,A_s r.
\label{eq:V-from-Asr}
\end{equation}
Accordingly, the characteristic inflationary energy scale may be written either in terms of observables or directly in terms of the WKB phase gradients:
\begin{equation}
V^{1/4}
\simeq
\left(\frac{3\pi^2}{2}A_s r\right)^{1/4}\Mpl
=
\left(\frac{\Pi_\alpha^2}{12\Mpl^2}\right)^{1/4}.
\label{eq:scale}
\end{equation}

Equations \eqref{eq:r-from-S}--\eqref{eq:scale} provide the explicit dictionary between the semiclassical wave function and the inflationary quantities constrained by observation.

\section{Semiclassical branch weights from the wave function}
\label{sec:weights}

To assign a semiclassical weight to an inflationary branch, one must specify the boundary condition that selects the wave function of the universe. In this context, the relevant compact saddle associated with the onset of inflation is Euclidean de Sitter space. Such a saddle exists only for closed spatial slices. The use of closed minisuperspace in Sec.~\ref{sec:minisuperspace} is therefore essential rather than merely conventional: it provides the minimal framework in which the semiclassical weighting of inflationary histories can be formulated in a well-defined way.

During the nucleation stage, we assume that the scalar field evolves sufficiently slowly that it may be approximated as constant across the Euclidean saddle,
\begin{equation}
\phi(\tau)\approx \phi_0,
\qquad
V(\phi)\approx V(\phi_0)\equiv V_0>0.
\label{eq:phi0}
\end{equation}
This is the standard adiabatic approximation employed in semiclassical analyses of inflationary nucleation \cite{Vilenkin1984,HawkingMoss1982,VilenkinYamada2018,Lehners2023}. Under this assumption, the saddle is governed by an effectively constant vacuum energy set by the local value of the scalar potential.

\subsection{Euclidean saddle}

In Euclidean signature, the action takes the form
\begin{equation}
S_E
=
-\int d^4x\,\sqrt{g}
\left[
\frac{\Mpl^2}{2}R - V_0
\right],
\label{eq:SE-def}
\end{equation}
where the scalar kinetic contribution has been neglected consistently with the approximation \eqref{eq:phi0}. On shell, the Euclidean Einstein equations imply
\begin{equation}
R_{\mu\nu} = \frac{V_0}{\Mpl^2} g_{\mu\nu},
\qquad
R = \frac{4V_0}{\Mpl^2}.
\label{eq:onshell-R}
\end{equation}
The corresponding saddle geometry is a four-sphere $S^4$ with radius
\begin{equation}
H_0^{-1},
\qquad
H_0^2 = \frac{V_0}{3\Mpl^2}.
\label{eq:H0}
\end{equation}
The volume of a four-sphere of radius $H_0^{-1}$ is
\begin{equation}
\mathrm{Vol}(S^4) = \frac{8\pi^2}{3H_0^4}.
\label{eq:S4vol}
\end{equation}

Substituting the on-shell relation \eqref{eq:onshell-R} into the Euclidean action \eqref{eq:SE-def}, one obtains
\begin{equation}
S_E
=
-\int d^4x\,\sqrt{g}
\left[
2V_0 - V_0
\right]
=
- V_0\,\mathrm{Vol}(S^4).
\end{equation}
Using \eqref{eq:S4vol} together with \eqref{eq:H0}, this becomes
\begin{align}
S_E
&=
- V_0 \frac{8\pi^2}{3}
\left(\frac{3\Mpl^2}{V_0}\right)^2
\nonumber\\[1mm]
&=
-\frac{24\pi^2 \Mpl^4}{V_0}.
\label{eq:SE-final}
\end{align}
This is the standard Euclidean de Sitter instanton action and provides the semiclassical exponent controlling the branch weight.

\subsection{Boundary-condition sign choice}

The sign with which the Euclidean action enters the semiclassical wave function is determined by the boundary condition. The two canonical choices are the Hartle-Hawking no-boundary prescription \cite{HartleHawking1983} and Vilenkin's tunneling prescription \cite{Vilenkin1982,Vilenkin1984,Vilenkin1988,VilenkinYamada2018}. It is convenient to write both cases in the unified form
\begin{equation}
\mathcal{W}_{\sigma}(\phi_0)
\propto
\exp[-\sigma S_E(\phi_0)]
=
\exp\!\left[
\sigma\,\frac{24\pi^2 \Mpl^4}{V(\phi_0)}
\right],
\label{eq:Wsigma}
\end{equation}
where
\begin{equation}
\sigma=
\begin{cases}
+1, & \text{no-boundary sign choice},\\[1mm]
-1, & \text{tunneling sign choice}.
\end{cases}
\label{eq:sigma-def}
\end{equation}
Equation \eqref{eq:Wsigma} gives the semiclassical branch weighting of inflationary initial data through the potential energy at the nucleation point.

At this stage, the conceptual separation between dynamics and weighting is fully explicit. The WKB phase determines the classical observables associated with a given branch, while the boundary condition fixes the relative weight assigned to different branches. The weighting formula \eqref{eq:Wsigma} does not modify the phase-to-observable correspondence derived in Sec.~\ref{sec:phase}, and the results of Sec.~\ref{sec:phase} do not determine the sign choice entering \eqref{eq:Wsigma}. These two ingredients play distinct and complementary roles within the semiclassical framework.

\subsection{A measure-independent threshold weight}

A fully normalized probability distribution over the space of inflationary initial conditions requires additional measure-theoretic input. For the present purposes, however, one can formulate sharp phenomenological statements without introducing such a measure.

In a single-field slow-roll model with monotonic evolution, let $\phi_N$ denote the field value for which the remaining duration of inflation is $N$ e-folds, as defined by
\begin{equation}
N
=
\frac{1}{\Mpl^2}
\int_{\phi_{\rm end}}^{\phi_N}
\frac{V(\phi)}{V_{,\phi}(\phi)}\,d\phi,
\label{eq:Nphi}
\end{equation}
where $\phi_{\rm end}$ marks the end of inflation. The corresponding quantity
\begin{equation}
\mathcal{W}_{\sigma}(N)
\equiv
\exp\!\left[
\sigma\,\frac{24\pi^2 \Mpl^4}{V(\phi_N)}
\right]
\label{eq:threshold-weight}
\end{equation}
may then be interpreted as the semiclassical weight of the threshold history that yields exactly $N$ e-folds. This object is not a normalized probability; rather, it is a measure-independent benchmark that isolates a physically meaningful question, namely how strongly a given boundary-condition sign weights the minimum initial energy density required to obtain a specified amount of inflation. Even in the absence of a global measure, this threshold weight already provides a sharp criterion for comparing boundary-condition sign choices in any explicit inflationary model.

\subsection{Controlled plateau limit}

An especially transparent limit arises when the potential remains sufficiently flat between the nucleation point and the epoch at which the pivot scale exits the Hubble radius, so that
\begin{equation}
V(\phi_0)\simeq V_*,
\label{eq:plateau-limit}
\end{equation}
with $V_*$ denoting the potential at horizon exit for the pivot mode. In this adiabatic plateau regime, the semiclassical branch weight can be written directly in terms of observable quantities. Using \eqref{eq:V-from-Asr}, one finds
\begin{equation}
\mathcal{W}_{\sigma}
\propto
\exp\!\left[
\sigma\,\frac{24\pi^2 \Mpl^4}{V_*}
\right]
=
\exp\!\left[
\sigma\,\frac{16}{A_s r}
\right].
\label{eq:W-obs}
\end{equation}
This relation is not exact for an arbitrary potential, since the approximation \eqref{eq:plateau-limit} need not hold universally. Whenever the plateau condition is satisfied, however, the weighting of inflationary branches becomes exponentially sensitive to the observable product $A_s r$, and therefore directly sensitive to the inflationary energy scale. In that regime, the connection between the wave function and phenomenology is especially direct.

\begin{figure}[t]
\centering
\includegraphics[width=0.82\textwidth]{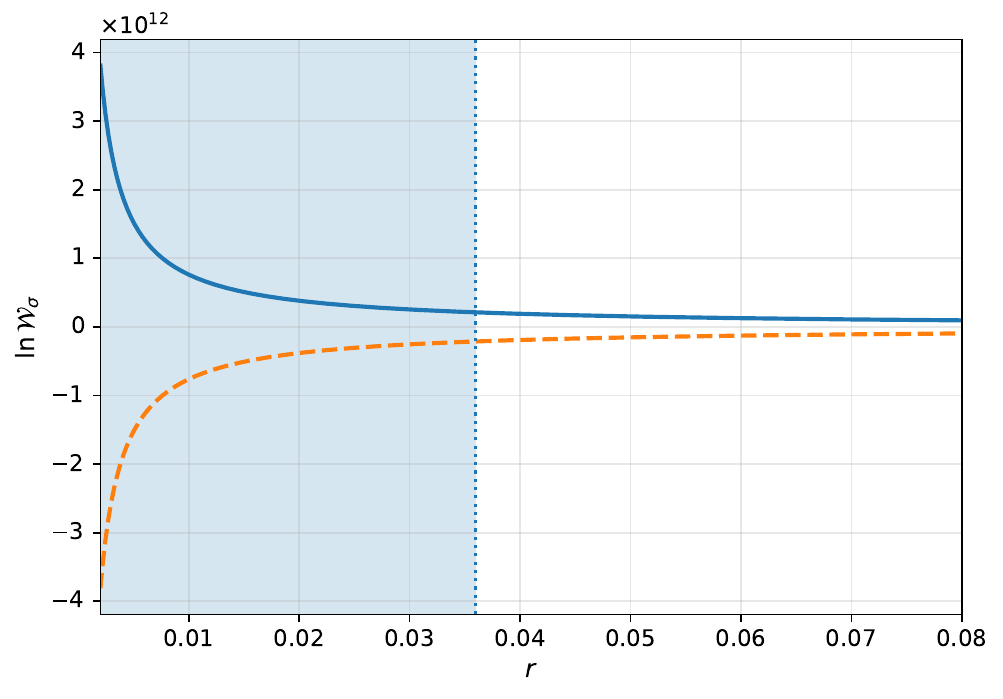}
\caption{Logarithm of the semiclassical branch weight, $\ln \mathcal{W}_{\sigma}$, as a function of the tensor-to-scalar ratio $r$ in the plateau limit, where $\ln \mathcal{W}_{\sigma}=\sigma\,16/(A_s r)$. The solid curve corresponds to $\sigma=+1$, and the dashed curve corresponds to $\sigma=-1$. The shaded region denotes the observationally allowed range implied by the current bound $r_{0.05}<0.036$, while the vertical dotted line marks the corresponding limiting value. The figure illustrates the strong inverse dependence of the semiclassical branch weight on the tensor amplitude once the scalar amplitude is fixed.}
\label{fig:lnWsigma_r}
\end{figure}

Figure~\ref{fig:lnWsigma_r} shows the plateau-limit dependence of the semiclassical branch weight on the tensor-to-scalar ratio. In this regime, the logarithm of the branch weight is given by
$\ln \mathcal{W}_{\sigma} = \sigma\,16/(A_s r)$,
so that the dependence on $r$ is explicitly inverse and therefore extremely steep at small tensor amplitude. The two curves correspond to the two sign choices $\sigma=\pm1$, while the shaded region indicates the range compatible with the current observational upper bound on $r$. The figure makes transparent that, once the scalar amplitude is fixed observationally, even modest changes in $r$ correspond to very large changes in the semiclassical weighting of inflationary branches.

\section{Phenomenological consequences}
\label{sec:phenomenology}

\subsection{Direct constraints on the WKB phase from current data}

Current cosmic microwave background constraints already translate into direct restrictions on the physically relevant phase gradients of the semiclassical wave function. Observations imply a scalar amplitude of approximately $A_s \simeq 2.1\times 10^{-9}$ and a scalar spectral tilt near $n_s \simeq 0.965$ from Planck, together with the upper bound $r_{0.05}<0.036$ from BICEP/Keck \cite{PlanckCosmo2018,PlanckInflation2018,BICEPKeck2021,BICEPKeckOverview2024}. Substituting this upper limit on $r$ into \eqref{eq:H-from-Asr} yields the bound
\begin{equation}
H_* < \pi \Mpl \sqrt{\frac{(2.1\times10^{-9})(0.036)}{2}}
\simeq 1.93\times 10^{-5}\,\Mpl.
\label{eq:Hbound}
\end{equation}
Equivalently, the corresponding upper bound on the inflationary energy scale is
\begin{equation}
V_*^{1/4}
<
\left[
\frac{3\pi^2}{2}(2.1\times 10^{-9})(0.036)
\right]^{1/4}\Mpl
\simeq 1.41\times10^{16}\,{\rm GeV},
\label{eq:Vbound}
\end{equation}
where we have used $\Mpl=2.435\times 10^{18}\,{\rm GeV}$.

These observational limits can be translated directly into constraints on the WKB phase gradients introduced in Sec.~\ref{sec:phase}. From \eqref{eq:Asr-from-S}, the geometric phase gradient at horizon exit satisfies
\begin{equation}
|\Pi_{\alpha *}|
=
\Mpl^3 \sqrt{18\pi^2 A_s r}
<
1.16\times 10^{-4}\,\Mpl^3.
\label{eq:Pialpha-bound}
\end{equation}
Similarly, \eqref{eq:r-from-S} implies the bound
\begin{equation}
\left|\frac{\Pi_{\phi *}}{\Pi_{\alpha *}}\right|
=
\sqrt{\frac{r}{288\,\Mpl^2}}
<
1.12\times 10^{-2}\,\Mpl^{-1}.
\label{eq:ratio-bound}
\end{equation}
Combining \eqref{eq:As-from-S} and \eqref{eq:r-from-S} then gives
\begin{equation}
|\Pi_{\phi *}|
=
\frac{\pi}{4}\,\sqrt{A_s}\,r\,\Mpl^2
<
1.30\times 10^{-6}\,\Mpl^2.
\label{eq:Piphi-bound}
\end{equation}

These relations provide the direct wave-function analogue of the observational bounds on the inflationary background. The geometric phase gradient is constrained by the upper limit on the Hubble scale, the scalar phase gradient is required to be smaller still because of the strong upper bound on primordial tensors, and the ratio of the two phase gradients is therefore tightly restricted. In this sense, the observed universe does not merely constrain an effective inflationary background; it constrains the local structure of the semiclassical phase of the wave function itself.

\subsection{What the scalar tilt constrains}

The observed scalar tilt constrains the directional variation of the WKB phase along the classical inflationary trajectory. Using \eqref{eq:ns-from-S}, one has
\begin{equation}
n_s-1
=
-36\,\Mpl^2\left(\frac{\Pi_\phi}{\Pi_\alpha}\right)^2
-
2\,D_N\ln\left|\frac{\Pi_\phi}{\Pi_\alpha}\right|.
\end{equation}
Since the measured spectrum is red, so that $n_s-1<0$, it follows that the combination
\begin{equation}
36\,\Mpl^2\left(\frac{\Pi_\phi}{\Pi_\alpha}\right)^2
+
2\,D_N\ln\left|\frac{\Pi_\phi}{\Pi_\alpha}\right|
\end{equation}
must be positive at horizon exit. The phenomenological content of this statement is clear. The scalar tilt is sensitive not only to the smallness of the scalar-to-geometric phase-gradient ratio, but also to the way in which that ratio evolves from one e-fold to the next. The observed red tilt therefore probes both the local orientation of the WKB phase in minisuperspace and its variation along the corresponding semiclassical branch.

\subsection{Threshold histories}

For any explicit inflationary potential, the threshold field value $\phi_N$ associated with a remaining duration of $N$ e-folds is determined by \eqref{eq:Nphi}. Inserting this field value into \eqref{eq:threshold-weight} gives
\begin{equation}
\mathcal{W}_{\sigma}(N)
=
\exp\!\left[
\sigma\,\frac{24\pi^2 \Mpl^4}{V(\phi_N)}
\right].
\end{equation}
This quantity is already sufficient to compare the two boundary-condition sign choices for the minimal inflationary history capable of producing $N$ e-folds.

The usefulness of this construction is twofold. First, it is fully explicit: the relevant semiclassical input is simply the value of the potential at the threshold configuration. Second, it is robust with respect to unresolved measure-theoretic issues: one does not need to specify a global probability measure on the full space of inflationary histories in order to compare how the two sign choices weight the threshold history required to satisfy phenomenological constraints. The threshold weight therefore provides a sharp and model-dependent diagnostic even in the absence of a complete measure prescription.

\subsection{Exponential sensitivity in the plateau limit}

In the controlled plateau regime defined by \eqref{eq:plateau-limit}, the branch weight depends directly on the observable product $A_s r$ through \eqref{eq:W-obs}. Using the measured scalar amplitude and the current upper bound on $r$, one finds
\begin{equation}
\frac{16}{A_s r}
>
\frac{16}{(2.1\times10^{-9})(0.036)}
\simeq 2.12\times 10^{11}.
\label{eq:huge-exponent}
\end{equation}
Accordingly, in the plateau regime the semiclassical branch weight is exponentially sensitive to the sign choice $\sigma$, with an exponent whose magnitude exceeds 
$ 2.12\times 10^{11} $
given the current observational bounds.

If the plateau approximation is valid, then the wave function assigns exponentially different weights to inflationary histories with different energy scales. The dependence on the observable combination $A_s r$ is therefore extraordinarily steep, and the semiclassical branch weighting retains a very strong sensitivity to the underlying inflationary scale.

\subsection{Summary of the phenomenological map}

The results of the preceding sections may be summarized as a direct correspondence between the semiclassical wave function and inflationary phenomenology. On a classical WKB branch, one has that the phase of the wave function determines the background kinematics and the associated inflationary observables. Separately, the choice of boundary condition fixes the relative weighting of semiclassical branches 
at the nucleation point. Taken together, these relations provide a compact and mathematically consistent framework for connecting the wave function of the universe to inflationary phenomenology.

\section{Conclusions}
\label{conclusions}

In this work, we developed a compact and explicit framework that connects the wave function of the universe to inflationary phenomenology within closed Friedmann-Robertson-Walker minisuperspace with a homogeneous scalar field. The analysis was organized around a clear separation between the dynamical information encoded in the semiclassical phase of the wave function and the probabilistic information encoded in the boundary condition. This separation provides a precise and physically transparent interpretation of how quantum cosmology enters the description of inflationary backgrounds.

On a classical WKB branch, the phase of the wave function plays the role of the Hamilton principal function, and its gradients determine the canonical momenta of the minisuperspace variables. This identification yields a direct correspondence between the phase gradients and the background quantities that govern inflationary evolution. In the regime relevant for observable inflation, where the initial curvature contribution has been diluted, this correspondence leads directly to expressions for the Hubble rate, the scalar-field velocity, the slow-roll hierarchy, the tensor-to-scalar ratio, the scalar amplitude, the scalar spectral tilt, and the inflationary energy scale in terms of the physical phase gradients of the wave function. The semiclassical phase therefore acquires an immediate phenomenological interpretation.

We also derived the semiclassical branch weights associated with standard quantum-cosmological boundary prescriptions by evaluating the Euclidean saddle for slowly varying scalar configurations. In this way, the weighting of inflationary histories is expressed directly in terms of the value of the scalar potential at the nucleation point. This result makes the role of the boundary condition fully explicit: it determines the relative weighting of semiclassical branches, while the subsequent inflationary kinematics remains governed by the WKB phase. The resulting framework thus assigns a precise and complementary role to each component of the wave function.

At the phenomenological level, the formalism already yields concrete consequences. Current cosmic microwave background data translate into direct constraints on the physical phase gradients associated with the semiclassical branch. The same framework provides a model-dependent threshold weight for obtaining a prescribed number of e-folds in any explicit slow-roll potential. In the controlled plateau regime, the branch weight can furthermore be written directly in terms of inflationary observables, making the dependence of the semiclassical weighting on the inflationary scale especially transparent.

Taken together, these results establish a direct and testable link between quantum cosmology and inflationary phenomenology. The central outcome of the present work is a two-part dictionary: the WKB phase determines the classical observables of an inflationary branch, and the boundary condition determines the relative weight assigned to that branch. Within this structure, the wave function of the universe acquires a clear phenomenological meaning and provides a mathematically coherent framework for relating quantum-cosmological initial conditions to observable properties of inflation.

\end{document}